\documentclass[useAMS,usenatbib]{mn2e}
\usepackage{graphicx}

\newcommand{\Teff}{\mbox{$T_{\mathrm{eff}}$}}
\newcommand{\Line}[3]{\Ion{#1}{#2}\,$\lambda$\,#3}
\newcommand{\Lines}[3]{\Ion{#1}{#2}\,$\lambda\lambda$\,#3}
\newcommand{\Ion}[2]{#1{\,\small #2}}

\newcommand{\Mwd}{\mbox{$M_{\mathrm{wd}}$}}
\newcommand{\Msun}{\mbox{$\mathrm{M}_{\odot}$}}

\title{SDSS\,J104341.53+085558.2: A second white dwarf with a
gaseous  debris disc}

\author[B.T. G\"ansicke et al.]{
B.T. G\"ansicke,
T.R. Marsh,
J. Southworth\\
Department of Physics, University of Warwick, Coventry CV4 7AL,
UK \\
}

\begin{document}

\date{Accepted 2005. Received 2005; in original form 2005}

\pagerange{\pageref{firstpage}--\pageref{lastpage}} \pubyear{2006}

\maketitle

\label{firstpage}

\begin{abstract}
Intermediate resolution spectroscopy of the white dwarf
SDSS\,J104341.53+085558.2 contains double-peaked emission lines of
\Lines{Ca}{II}{8498,8542,8662} and identifies this object to be the
second single white dwarf to be surrounded by a gaseous disc of
metal-rich material, similar to the recently discovered
SDSS\,J1228+1040. A photospheric Magnesium abundance of 0.3 times the
solar value, determined from the observed \Line{Mg}{II}{4481}
absorption line, implies that the white dwarf is accreting from the
circumstellar material. The absence of Balmer emission lines and of
photospheric \Line{He}{I}{4471} absorption indicates that the accreted
material is depleted in volatile elements and, by analogy with
SDSS\,1228+1040, may be the result of the tidal disruption of an
asteroid.  Additional spectroscopy of the DAZ white dwarfs
WD\,1337+705 and GD362 does not reveal \Ion{Ca}{II} emission
lines. GD362 is one of the few cool DAZ that display strong infrared
flux excess, thought to be originating in a circumstellar dust disc,
and its temperature is likely too low to sublimate sufficient amounts
of disc material to generate detectable \Ion{Ca}{II}
emission. WD\,1337+705 is, as SDSS\,1228+1040 and SDSS\,J1043+0855,
moderately hot, but has the lowest Mg abundance of those three stars,
suggesting a possible correlation between the photospheric Mg
abundance and the equivalent width of the \Ion{Ca\,II} emission
triplet. Our inspection of 7360 white dwarfs from SDSS DR\,4
fails to unveil additional strong ``metal gas disc'' candidates, and
implies that these objects are rather rare.
\end{abstract}


\section{Introduction}
Two decades ago, \citet{zuckerman+becklin87-1} detected an infrared
excess around the DA white dwarf G29--38 as part of their search for
cool companions to white dwarfs. \citet{grahametal90-2} and
\citet{kuchneretal98-1} convincingly ruled out the presence of a
low-mass stellar companion, and \citet{grahametal90-1} suggested that
the observed infrared excess is caused by a ring of circumstellar dust
around the white dwarf, a model further developed by
\citet{jura03-1,jura06-1}, and underpinned by Spitzer observations
\citep{reachetal05-1}. Another four single white dwarfs with infrared
excesses have since then been discovered (GD362,
\citealt{becklinetal05-1, kilicetal05-1}; GD56,
\citealt{kilicetal06-1}; WD\,1150-153, \citealt{kilic+redfield07-1};
WD\,2115--560, von Hippel et al. 2007), plus one additional candidate
(G167--8, \citealt{farihietal06-1}). Two common characteristic of all
those white dwarfs with circumstellar dust discs are their low
temperatures, $\Teff<15000$\,K, and their substantial abundances of
photospheric Calcium \citep{koesteretal97-1, zuckermanetal03-1,
koesteretal05-2}. No evidence for the presence of dusty discs has been
found around white dwarfs hotter than $\simeq16\,000$\,K
\citep{kilicetal06-1}.

\begin{figure*}
\centerline{\includegraphics[angle=-90,width=13cm]{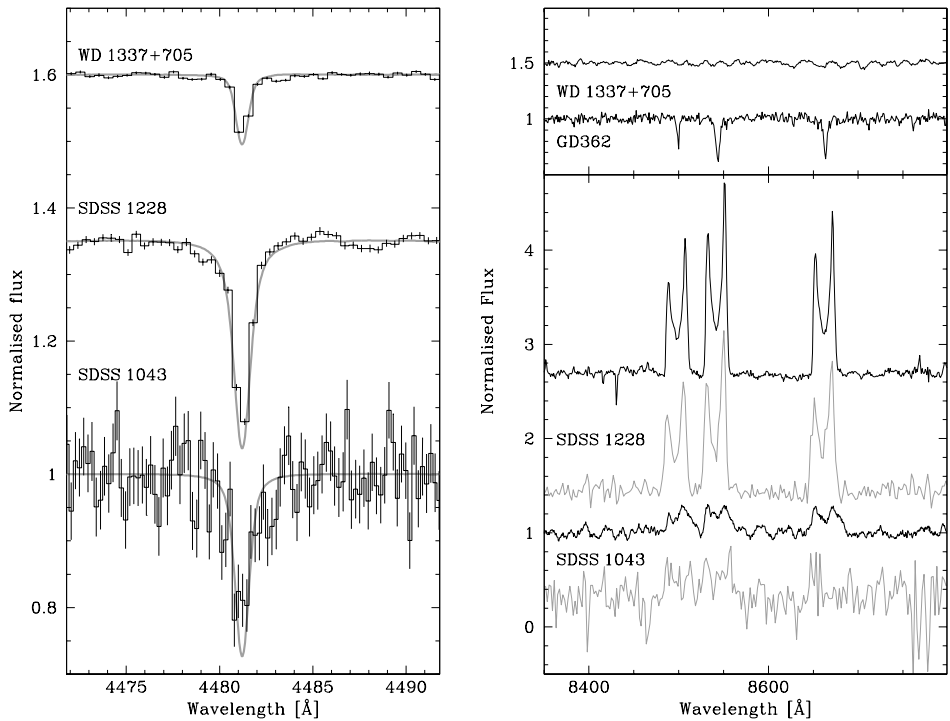}}
\caption{\label{f-caii} Left panel: photospheric \Line{Mg}{II}{4481}
absorption lines in the WHT spectra of WD\,1337+705, SDSS\,1228+1040,
and SDSS\,1043+0855 (black lines). Overplotted in gray are the
best-fit white dwarf models, the corresponding Mg abundances are given in
Table\,\ref{t-abundances}. Right panel: WHT (black lines) and SDSS
(gray lines) spectra of SDSS\,1043+0855, SDSS\,1228+1040, WD\,1337+705
and GD362. All spectra are normalised to a continuum flux of one, and
offset by suitable amounts. The top panel shows the WHT spectra of 
WD\,1337+705 and GD362 on a different flux scale.}
\end{figure*}

Recently, \citet{gaensickeetal06-3} detected emission lines of
\Lines{Ca}{II}{8498,8542,8662} in the DA white dwarf
SDSS\,J122859.93+104032.9, which has a temperature of 22\,000\,K, much
hotter than the five white dwarfs exhibiting infrared excess. The
double-peaked shape of the \Ion{Ca}{II} emission lines unambiguously
identifies the presence of a rotating ring of gas around
SDSS\,1228+1040.  Time-resolved spectroscopic and photometric follow-up
observations of SDSS\,1228+1040 ruled out the possibility of it being
a close binary system where an accretion disc would form from material
lost by the companion star.  

The detection of a strong \Line{Mg}{II}{4481} absorption line
identifies SDSS\,1228+1040 as a DAZ white dwarf, and indicates a
photospheric Magnesium abundance close to the solar value.  Given that
the gravitational sedimentation time scales in the radiative
atmosphere are very short \citep{koester+wilken06-1} and radiative
levitation is negligible \citep{chayeretal94-1}, it is clear that
SDSS1228+1040 is accreting from the circumstellar gas disc. 

The absence of hydrogen or helium emission lines from the ring, along
with the absence of helium absorption lines from the photosphere of
the white dwarf, indicates that the circumstellar disc must be
depleted in volatile elements, and \citet{gaensickeetal06-3} concluded
that the most likely origin of this disc is a tidally disrupted
asteroid. SDSS\,1228+1040 with its circumstellar gas disc appears
hence as the hot counterpart to G29--38 and the other cool DAZ
harbouring dust discs.  \citet{gaensickeetal06-3} identified
SDSS\,J104341.53+085558.2 (henceforth SDSS\,1043+0855) as a good
candidate for being the second white dwarf with a circumstellar
gaseous metal disc. We present here follow-up observations that
confirm this suggestion.

\section{Observations}
Intermediate resolution spectroscopy of SDSS\,1043+0855 was obtained
at the William Herschel Telescope (WHT) in service mode on 2007,
February 3, using the double-arm spectrograph ISIS. The blue arm was
equipped with the R1200B grating and a 4k$\times$2k pixel EEV
detector, providing a spectral coverage of $\simeq4000-4700$\,\AA\ at
a resolution (FWHM) of $\simeq0.9$\,\AA. In the red arm, the R600R
grating was used along with the low-fringing 4k$\times$2k pixel
REDPLUS detector, providing a spectral coverage of
$\simeq7460-9100$\,\AA\ at a resolution of $\simeq2$\,\AA. A total of
4 pairs of blue/red spectra with individual exposure times of 20\,min
were obtained under poor seeing ($\sim2.5"$) conditions. In June 2006,
we also obtained WHT spectroscopy of GD\,362, which is one of the cool
($\Teff=9740$\,K, \citealt{gianninasetal04-1}) DAZ white dwarfs
exhibiting infrared excess \citep{becklinetal05-1, kilicetal05-1}, and
the hotter DAZ WD\,1337+705 (Grw+70 5824), with an identical setup,
except for using the 4.5k$\times$2k pixel Marconi detector in the red
arm, which is subject to noticeable fringing in the red end of the
spectrum. All data were reduced in a standard way using
\texttt{STARLINK} software and the \texttt{Pamela/Molly} packages.

As anticipated from its SDSS spectrum, SDSS\,1043+0855 displays
double-peaked \Lines{Ca}{II}{8498,8542,8662} lines
(Fig.\,\ref{f-caii}, right panel), though at substantially lower
strengths compared to SDSS\,1228+1040. The relatively low quality of
the $I$-band spectra prevents a dynamical analysis of the disc
emission, but the morphology of the line profiles in SDSS\,1043+0855
is fairly similar to those observed in SDSS\,1228+1040
\citep{gaensickeetal06-3}, suggesting broadly similar parameters.  No
significant trace of \Ion{Ca}{II} emission is found in WD\,1337+705, within
the limitations imposed by the CCD fringing. The red spectrum of
GD\,362 contains  \Ion{Ca}{II} in absorption from the white dwarf
photosphere. The spectra from the blue arm reveal the presence of
photospheric \Line{Mg}{II}{4481} absorption in SDSS\,1043+0855 and
WD\,1337+705 (Fig.\,\ref{f-caii}, left panel). GD\,362 is too cold
\citep{gianninasetal04-1} to exhibit significant \Line{Mg}{II}{4481}
absorption. The absence of Zeeman splitting in the Ca triplet limits
the magnetic field strength in GD362 to $\la 30$kG (see
\citealt{dufouretal06-1} for the weakly magnetic DZ G165--7).

\begin{figure*}
\includegraphics[angle=-90,width=18cm]{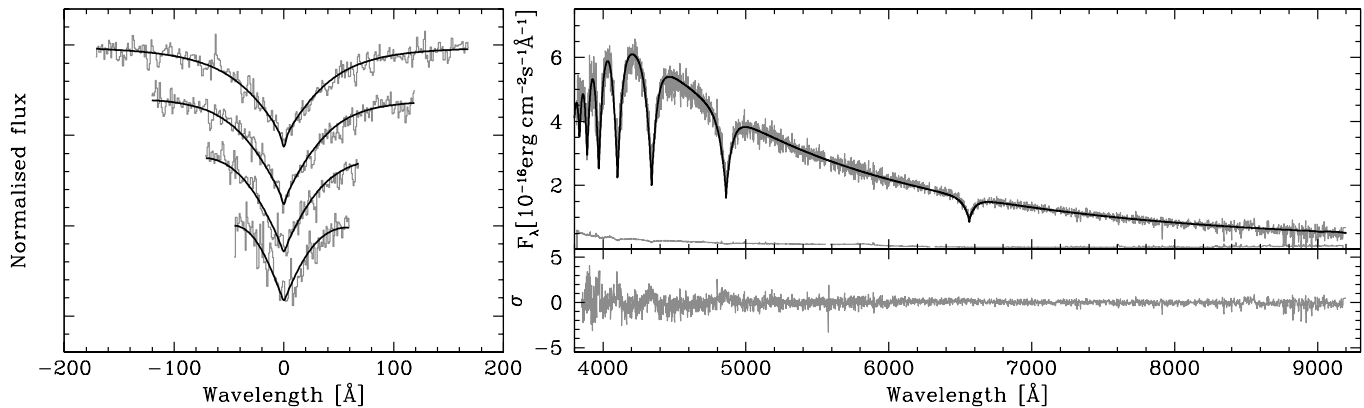}
\caption{\label{f-opt_fit} Top right panel: the SDSS spectrum and
    flux error of SDSS\,1043+0855 (gray lines; plate 1240, MJD 52734,
    fibre 37), along with the best fit to the full spectral range
    (black line), which is dominated by the slope of the continuum
    (hence the significant residuals in the Balmer lines, shown in the
    bottom right panel).  Left
    panel: the best fit (black lines) to the to the normalised
    H$\beta$ to H$\epsilon$ line profiles (gray lines, top to bottom),
    which is used to determine the effective temperature and surface
    gravity ($\Teff=18\,330$\,K, $\log g=8.09$).}
\end{figure*}

\section{White dwarf parameter of SDSS1043+0855}
We have used a grid of model spectra calculated with TLUSTY/SYNSPEC
\citep{hubeny88-1,hubeny+lanz95-1} to analyse the SDSS and WHT spectra
of SDSS\,1043+0855. The model atmospheres were computed assuming a
pure hydrogen composition and local thermodynamic equilibrium (LTE), and
sequences of synthetic spectra were subsequently calculated for a
variety of Mg abundances. In order to determine the temperature
and surface gravity of the white dwarf, we fitted both the entire
spectrum, as well as the normalised H$\alpha$ to H$\epsilon$ lines.
The Balmer lines reach their maximum equivalent widths around
$\Teff=13\,500$\,K for $\log g=8$, or $\sim1000$\,K higher (lower) for
$\log g=8.5$ $(\log g=7.5)$, and consequently a fit to the normalised
Balmer line profiles results usually in a ``hot'' and a ``cold''
solution of comparable quality.  We use the fit to the entire data,
continuum plus lines, to choose the solution which better agrees with
the slope of the spectrum. Our fit includes H$\alpha$--H$\epsilon$.
The higher Balmer lines, even though more sensitive to the surface
gravity \citep[e.g.][]{kepleretal06-1}, are of too poor a quality in
the SDSS data to be useful. The best-fit $\Teff$ and $\log g$ and
their 1-$\sigma$ errors are obtained from a bicubic spline
interpolation to the $\chi^2$ values on the $\Teff-\log g$ grid
covered by our model spectra.

The parameters from the best fit to the normalised line profiles are
$\Teff=18330\pm523$\,K and $\log g=8.09\pm0.11$
(Fig.\,\ref{f-opt_fit}). Using an updated version of the evolutionary
sequences in \citet{bergeronetal95-2}, a white dwarf mass of
$\Mwd=0.67\pm0.07\,\Msun$, a radius of $8.55\pm0.65\times10^8$\,cm as
well as a cooling age of $1.3\times10^8$\,y are derived. The flux
scaling factor between the observed and model fluxes implies a
distance of $224\pm18$\,pc.  While we find good agreement with the
surface gravity determined by \citet{eisensteinetal06-1} with their
\texttt{autofit} procedure, $\log g=8.06\pm0.07$. Our effective
temperature is hotter by $\sim1300$\,K compared to their value of
$\Teff=17044\pm288$, suggesting that differing details in the fitting
procedure cause systematic uncertainties that can be somewhat larger
than the statistical errors.

\vspace*{-3ex}
\section{Photospheric \Ion{Mg}{II} abundances and \Ion{Ca}{II} emission line
  equivalent widhts}

We have determined the photospheric Mg abundances of SDSS\,1043+0855
and WD\,1337+305 by fitting TLUSTY/SYNSPEC models with the best-fit
$\Teff$ and $\log g$ but variable Mg abundances to the normalised
\Line{Mg}{II}{4481} line profile observed in the WHT spectra. The
widths of the observed \Ion{Mg}{II} lines are consistent in both cases with
very low rotational velocities, $v\sin i<15\mathrm{km\,s^{-1}}$. We
find for SDSS\,1043+0855 a Mg abundances of $0.30\pm0.15$ times the
solar value, or $\log(\mathrm{Mg/H})=-4.94\pm{0.17\atop0.30}$. For
WD\,1337+705, our fit results in a Mg abundance of $0.07\pm0.01$ times
the solar value, or $\log(\mathrm{Mg/H})=-5.58\pm{0.06\atop0.06}$,
which is in good agreement with the measurement of
\citet{zuckermanetal03-1}. Both our and Zuckerman's Mg abundance
measurements are somewhat lower than that of \citet{holbergetal97-1},
$\log(\mathrm{Mg/H})=-5.35\pm0.10$, which was determined from a rather
noisy spectrum.

\begin{figure}
\centerline{\includegraphics[width=6cm]{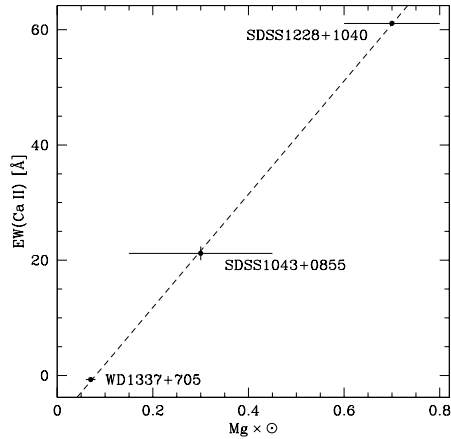}}
\caption{\label{f-mgline} Correlation of the Mg abundance in the white
dwarf photosphere and the equivalent width of the
\Lines{Ca}{II}{8498,8542,8662} triplet in the three hot DAZ observed
by us in the $I$-band. }
\end{figure}

We have measured the equivalent widths of the combined \Ion{Ca}{II} triplet
in out WHT spectra of SDSS\,1043+0855, SDSS\,1228+1040 (from
\citealt{gaensickeetal06-3}), and WD\,1337+705. For WD\,1337+705, the
WHT spectrum is consistent with no \Ion{Ca}{II} emission at all
(Table\,\ref{t-abundances}). Figure\,\ref{f-mgline} shows the
correlation between the photospheric Mg abundances in SDSS\,1043+0855,
SDSS\,1228+1040, and WD\,1337+705 and the equivalent widths of the
Ca\,II triplet.

\begin{table}
\caption{\label{t-abundances} Photospheric Mg abundances and the
combined equivalent width of the \Lines{Ca}{II}{8498,8542,8662}
triplet in SDSS\,1043+0855, WD\,1337+705, and SDSS\,1228+1040.}
\begin{tabular}{lccc}
\hline
Object & Mg $\times(\odot)$ & $\log(\mathrm{Mg/H})$ & EW(Ca\,II) [\AA]\\
\hline
SDSS\,1043+0855 & $0.30\pm0.15$ & $-4.94\pm{0.17\atop0.30}$ & $21.2\pm1.2$\\
WD\,1337+705    & $0.07\pm0.01$ & $-5.58\pm{0.06\atop0.06}$ & $-0.7\pm0.1$\\ 
SDSS\,1228+1040 & $0.70\pm0.10$ & $-4.58\pm{0.06\atop0.06}$ & $61.1\pm0.2$\\
\hline
\end{tabular}
\end{table}

\vspace*{-3ex}
\section{More SDSS\,1228+1040 stars in SDSS\,DR4?}
\citet{gaensickeetal06-3} visually inspected the spectra of
406 DA white dwarfs brighter than $g=17.5$ from DR4, and the only
viable candidate for the presence of \Ion{Ca}{II} emission lines was
SDSS\,1043+0855. In order to put a more quantitative constraint on the
number of white dwarfs with circumstellar discs of metal-rich gas,
we implemented an automated measurement of the
\Lines{Ca}{II}{8498,8542,8662} triplet in \texttt{SuperMongo}. In
brief, this routine extracts and normalises the white dwarf spectrum
in the wavelength range 8000--9200\,\AA, dividing it by a first-order
polynomial fit to the line-free continuum (8000--8450\,\AA\ and
8725--9000\,\AA). In a second step, the combined equivalent width of
the \Ion{Ca}{II} triplet is calculated by integrating the normalised
spectrum the range 8465--8690\,\AA, dividing by the bandwidth of this
interval, and subtracting the corresponding continuum
contribution. The flux errors of the spectrum are propagated in an
equivalent fashion.  We downloaded the SDSS spectra of all DA white
dwarfs from the \citet{eisensteinetal06-1} list, only excluding those
classified as DA+K binaries (but including those with an uncertain
binary classification, DA+K:), resulting in a total of 7360 individual
objects. Subjecting those spectra to the procedure outlined above
produced a list of 300 white dwarfs with a $3\sigma$ excess in the
\Ion{Ca}{II} triplet over the neighbouring continuum. The SDSS spectra of
these objects were then visually inspected.

Both SDSS\,1228+1040 and SDSS\,1043+0855 were recovered by this
automated search as the two most obvious Ca\,II emssion line
candidates. The vast majority of additional candidates turned out to
be faint ($\sim i>19$) white dwarfs with substantial residuals from
the night sky subtraction.  A total of 8 additional rather
weak candidates for \Ion{Ca}{II} emission were identified and are
listed in Table\,\ref{t-candidates}. While this exercise
confirms the finding of \citet{gaensickeetal06-3} that white dwarfs
with gaseous metal discs are rare, it also shows that the SDSS data is
of sufficient quality only for the brightest $\sim2000$ DA white
dwarfs from Eisenstein's (\citeyear{eisensteinetal06-1}) list, a proper
assessment of white dwarfs fainter than $i\sim19$ will require better
data, with a particular emphasis on a good skyline subtraction.

\begin{table}
\caption{\label{t-candidates} Confirmed (in italics) and candidate white dwarfs
  with \Lines{Ca}{II}{8498,8542,8662} emission. $\Teff$ and $\log g$
  are determined from fitting the SDSS spectra as described in
  Sect.\,3. The combined equivalent width (EW) for the triplet is
  given.}
\setlength{\tabcolsep}{0.8ex}
\begin{tabular}{lcrcrrr}
\hline
SDSS\,J & $g$ & \Teff[K] & $\log g$ & EW[\AA] & Notes\\
\hline
015854.17+123813.3 & 18.1 & $7910\pm30$ & $8.45\pm0.10$ & 8.1 & 1 \\
023543.07+005557.0 & 18.4 & $10307\pm142$ & $8.58\pm0.15$ & 19.4 & 1,2 \\
075409.24+485058.1 & 19.4 & $35154\pm1877$ & $7.70\pm0.33$ & 134 &   \\
090555.02+034006.3 & 19.0 & $16149\pm576$  & $7.99\pm0.13$ & 28.4 & \\
093956.34+390712.2 & 19.4 & $9081\pm209$ & $8.43\pm0.35$ & 41.6  & 1 \\
\textit{104341.53+085558.2} & 17.5 & $18330\pm523$ & $8.09\pm0.11$ & 21.2 & 3 \\
111701.96+000322.9 & 19.2 & $20804\pm807$ & $8.07\pm0.15$ & 33.6 & 2\\
\textit{122859.93+104032.9} & 16.7 & $22292\pm296$ & $8.29\pm0.05$ & 61.1 & 4 \\
144849.62+024024.9 & 17.7 & $15246\pm315$ & $7.50\pm0.08$ & 13.9 & \\
224753.21-000230.2 & 19.0 & $7641\pm308$ & $8.73\pm0.72$ & 18.4 & 1 \\
\hline
\end{tabular}
\begin{minipage}{\columnwidth}
$^1$ Too cold for sublimating circumstellar material at the tidal
  destruction radius of the white dwarf \citep[e.g.][]{jura03-1}.
$^2$ More than one SDSS spectrum available, excess clearly visible
  only in one of them. 
$^3$ This paper.
$^4$ The white dwarf parameter differ very slightly with respect to
 those in \citet{gaensickeetal06-3} due to improvments in our fitting procedure.
\end{minipage}
\end{table}

\vspace*{-3ex}
\section{Discussion and Conclusions}
The discovery of gaseous discs around the two moderately hot
white dwarfs SDSS\,1228+1040 and SDSS\,1043+0855 and dust discs around
white dwarfs with $\Teff\la15\,000$\,K indicates that the white dwarf
temperature plays a crucial role in determing the phase state of
circumstellar debris discs. Von Hippel et
al. (\citeyear{vonhippeletal07-1}) explored the range of white dwarf
effective temperatures for which the sublimation radius is inside of
the Roche radius for tidal disruption, and found good agreement with
the observational evidence.

The origin of metals in the photospheres of white dwarfs has been
intensively debated (see \citealt{koester+wilken06-1,
kilic+redfield07-1, vonhippeletal07-1}). A purely interstellar origin,
as e.g. worked out in detail by \citet{dupuisetal93-1}, appears less
likely in the view of the observations collected throughout the past
15 years. The detection of dusty \citep[e.g.][]{zuckerman+becklin87-1,
reachetal05-1, becklinetal05-1, kilicetal05-1, vonhippeletal07-1} and
gaseous discs (\citealt{gaensickeetal06-3} and this paper) of hydrogen
and helium depleted material offers a viable alternative at least for
some systems: accretion from tidally disrupted asteroids
\citep{jura03-1}. However, only a relatively small fraction of the
cool ($\Teff\la15\,000$\,K) DAZ white dwarfs exhibit infrared excess,
and it is currently not clear if the photospheric metals found in the
remaining systems is also associated with the presence of planetary
debris. \citet{vonhippeletal07-1} show that the 5 confirmed white
dwarfs with dusty discs have accretion rates at the upper end of what
is observed in cool DAZ, and our observations give some evidence that
the strength of the \Ion{Ca}{II} emission correlates with the the
photospheric Mg abundance (Fig.\,\ref{f-mgline}) . It may hence be
that the white dwarfs with clearly visible discs represent only the
``tip of the iceberg''. 

The \Ion{Ca}{II} emission lines detected in SDSS\,1228+1040 and
SDSS\,1043+0855 offer substantialy dynamical insight into the structures
of the circumstellar discs, and long-term monitoring of these line profiles
appears worthwhile to probe for evolution of the disc radii and
eccentricities.

\vspace*{-3ex}
\section*{Acknowledgements}
JS was supported by a PPARC PDRA. 
Based on observations made with the William Herschel Telescope, which
is operated on the island of La Palma by the Isaac Newton Group in the
Spanish Observatorio del Roque de los Muchachos of the Instituto de
Astrof{\'\i}sica de Canarias.
We thank the referee for a constructive report, and 
Ben Zuckerman for his comments on the submitted manuscript.

\section*{Note added in proof}
After the submission of this paper, \citet{juraetal07-1}
reported the Spitzer detection of an infrared excess for  white dwarf
PG\,1015+161 ($\Teff=19\,300$\,K) suggesting that gas and dust discs
may co-exist over a certain range in white dwarf effective temperature.


\begin{thebibliography}{29}
\expandafter\ifx\csname natexlab\endcsname\relax\def\natexlab#1{#1}\fi

\bibitem[{{Becklin} et~al.(2005){Becklin}, {Farihi}, {Jura}, {Song},
  {Weinberger}, \& {Zuckerman}}]{becklinetal05-1}
{Becklin}, E.~E., {Farihi}, J., {Jura}, M., {Song}, I., {Weinberger}, A.~J.,
  {Zuckerman}, B., 2005, ApJ Lett., 632, L119

\bibitem[{{Bergeron} et~al.(1995){Bergeron}, {Wesemael}, \&
  {Beauchamp}}]{bergeronetal95-2}
{Bergeron}, P., {Wesemael}, F., {Beauchamp}, A., 1995, PASP, 107, 1047

\bibitem[{{Chayer} et~al.(1994){Chayer}, {Leblanc}, {Fontaine}, {Wesemael},
  {Michaud}, \& {Vennes}}]{chayeretal94-1}
{Chayer}, P., {Leblanc}, F., {Fontaine}, G., {Wesemael}, F., {Michaud}, G.,
  {Vennes}, S., 1994, ApJ Lett., 436, L161

\bibitem[{{Dufour} et~al.(2006){Dufour}, {Bergeron}, {Schmidt}, {Liebert},
  {Harris}, {Knapp}, {Anderson}, \& {Schneider}}]{dufouretal06-1}
{Dufour}, P., {Bergeron}, P., {Schmidt}, G.~D., {Liebert}, J., {Harris}, H.~C.,
  {Knapp}, G.~R., {Anderson}, S.~F., {Schneider}, D.~P., 2006, ApJ, 651, 1112

\bibitem[{{Dupuis} et~al.(1993){Dupuis}, {Fontaine}, {Pelletier}, \&
  {Wesemael}}]{dupuisetal93-1}
{Dupuis}, J., {Fontaine}, G., {Pelletier}, C., {Wesemael}, F., 1993, ApJS, 84,
  73

\bibitem[{{Eisenstein} et~al.(2006)}]{eisensteinetal06-1}
{Eisenstein}, D.~J., et~al., 2006, ApJS, 167, 40

\bibitem[{{Farihi} et~al.(2007){Farihi}, {Zuckerman}, {Becklin}, \&
  {Jura}}]{farihietal06-1}
{Farihi}, J., {Zuckerman}, B., {Becklin}, E.~E., {Jura}, M., 2007, in
  {Napiwotzki}, R., {Burleigh}, R., eds., 15th European Workshop on White
  Dwarfs, ASP Conf. Ser., p. in press

\bibitem[{{G{\"a}nsicke} et~al.(2006){G{\"a}nsicke}, {Marsh}, {Southworth}, \&
  {Rebassa-Mansergas}}]{gaensickeetal06-3}
{G{\"a}nsicke}, B.~T., {Marsh}, T.~R., {Southworth}, J., {Rebassa-Mansergas},
  A., 2006, Science, 314, 1908

\bibitem[{{Gianninas} et~al.(2004){Gianninas}, {Dufour}, \&
  {Bergeron}}]{gianninasetal04-1}
{Gianninas}, A., {Dufour}, P., {Bergeron}, P., 2004, ApJ Lett., 617, L57

\bibitem[{{Graham} et~al.(1990{\natexlab{a}}){Graham}, {Matthews},
  {Neugebauer}, \& {Soifer}}]{grahametal90-1}
{Graham}, J.~R., {Matthews}, K., {Neugebauer}, G., {Soifer}, B.~T.,
  1990{\natexlab{a}}, ApJ, 357, 216

\bibitem[{{Graham} et~al.(1990{\natexlab{b}}){Graham}, {Reid}, {McCarthy}, \&
  {Rich}}]{grahametal90-2}
{Graham}, J.~R., {Reid}, I.~N., {McCarthy}, J.~K., {Rich}, R.~M.,
  1990{\natexlab{b}}, ApJ Lett., 357, L21

\bibitem[{{Holberg} et~al.(1997){Holberg}, {Barstow}, \&
  {Green}}]{holbergetal97-1}
{Holberg}, J.~B., {Barstow}, M.~A., {Green}, E.~M., 1997, ApJ Lett., 474, L127

\bibitem[{{Hubeny}(1988)}]{hubeny88-1}
{Hubeny}, I., 1988, Comput.,Phys.,Comm., 52, 103

\bibitem[{{Hubeny} \& {Lanz}(1995)}]{hubeny+lanz95-1}
{Hubeny}, I., {Lanz}, T., 1995, ApJ, 439, 875

\bibitem[{{Jura}(2003)}]{jura03-1}
{Jura}, M., 2003, ApJ Lett., 584, L91

\bibitem[{{Jura}(2006)}]{jura06-1}
{Jura}, M., 2006, ApJ, 653, 613

\bibitem[{{Jura} et~al.(2007){Jura}, {Farihi}, \& {Zuckerman}}]{juraetal07-1}
{Jura}, M., {Farihi}, J., {Zuckerman}, B., 2007, ApJ, in press, arxiv:0704.1170

\bibitem[{{Kepler} et~al.(2006){Kepler}, {Castanheira}, {Costa}, \&
  {Koester}}]{kepleretal06-1}
{Kepler}, S.~O., {Castanheira}, B.~G., {Costa}, A.~F.~M., {Koester}, D., 2006,
  MNRAS, 372, 1799

\bibitem[{{Kilic} \& {Redfield}(2007)}]{kilic+redfield07-1}
{Kilic}, M., {Redfield}, S., 2007, ApJ, 660, 641

\bibitem[{{Kilic} et~al.(2005){Kilic}, {von Hippel}, {Leggett}, \&
  {Winget}}]{kilicetal05-1}
{Kilic}, M., {von Hippel}, T., {Leggett}, S.~K., {Winget}, D.~E., 2005, ApJ
  Lett., 632, L115

\bibitem[{{Kilic} et~al.(2006){Kilic}, {von Hippel}, {Leggett}, \&
  {Winget}}]{kilicetal06-1}
{Kilic}, M., {von Hippel}, T., {Leggett}, S.~K., {Winget}, D.~E., 2006, ApJ,
  646, 474

\bibitem[{{Koester} \& {Wilken}(2006)}]{koester+wilken06-1}
{Koester}, D., {Wilken}, D., 2006, A\&A, 453, 1051

\bibitem[{{Koester} et~al.(1997){Koester}, {Provencal}, \&
  {Shipman}}]{koesteretal97-1}
{Koester}, D., {Provencal}, J., {Shipman}, H.~L., 1997, A\&A, 320, L57

\bibitem[{{Koester} et~al.(2005){Koester}, {Rollenhagen}, {Napiwotzki}, {Voss},
  {Christlieb}, {Homeier}, \& {Reimers}}]{koesteretal05-2}
{Koester}, D., {Rollenhagen}, K., {Napiwotzki}, R., {Voss}, B., {Christlieb},
  N., {Homeier}, D., {Reimers}, D., 2005, A\&A, 432, 1025

\bibitem[{{Kuchner} et~al.(1998){Kuchner}, {Koresko}, \&
  {Brown}}]{kuchneretal98-1}
{Kuchner}, M.~J., {Koresko}, C.~D., {Brown}, M.~E., 1998, ApJ Lett., 508, L81

\bibitem[{{Reach} et~al.(2005){Reach}, {Kuchner}, {von Hippel}, {Burrows},
  {Mullally}, {Kilic}, \& {Winget}}]{reachetal05-1}
{Reach}, W.~T., {Kuchner}, M.~J., {von Hippel}, T., {Burrows}, A., {Mullally},
  F., {Kilic}, M., {Winget}, D.~E., 2005, ApJ Lett., 635, L161

\bibitem[{{von Hippel} et~al.(2007){von Hippel}, {Kuchner}, {Kilic},
  {Mullally}, \& {Reach}}]{vonhippeletal07-1}
{von Hippel}, T., {Kuchner}, M.~J., {Kilic}, M., {Mullally}, F., {Reach},
  W.~T., 2007, ApJ, in press, astro-ph/0703473

\bibitem[{{Zuckerman} \& {Becklin}(1987)}]{zuckerman+becklin87-1}
{Zuckerman}, B., {Becklin}, E.~E., 1987, Nat, 330, 138

\bibitem[{{Zuckerman} et~al.(2003){Zuckerman}, {Koester}, {Reid}, \&
  {H{\"u}nsch}}]{zuckermanetal03-1}
{Zuckerman}, B., {Koester}, D., {Reid}, I.~N., {H{\"u}nsch}, M., 2003, ApJ,
  596, 477

\end{thebibliography}

\bsp

\label{lastpage}

\end{document}